\documentclass{article}

\usepackage{graphicx}
\usepackage[all]{xypic}
\usepackage{eurosym}
\usepackage{amsmath}
\usepackage{amsfonts}
\usepackage{amssymb}
\usepackage{amsthm}
\usepackage{mathrsfs}
\usepackage{newlfont}
\usepackage{hyperref}
\usepackage{times}
\usepackage{bm}
\usepackage[figurename=Fig.]{caption}
\usepackage{longtable}

\numberwithin{equation}{section}

\newcommand{\abs}[1]{\left\vert#1\right\vert}

\newcommand{\To}{\longrightarrow}

\begin{document}

\title{\textsc{Strong Field of Binary Systems And Its Effects On Pulsar Arrival Times}}%

\author{M.I. Wanas\footnote{Astronomy department, faculty of science,
Cairo university, Egypt. Center for theoretical physics at the British university in Egypt. e-mail: wanas@cu.edu.eg; mamdouh.wanas@bue.edu.eg}, N.S. Awadalla\footnote{National research institute for astronomy and geophysics, Egypt. e-mail: awadalla@frcu.eun.eg} and W.S. El Hanafy\footnote{The British University in Egypt, Faculty of Engineering, Basic Science Department. Centre for theoretical physics at the British University in Egypt, 11837 - P.O. Box 43, Egypt. e-mail: waleed.elhanafy@bue.edu.eg}}
\maketitle
\begin{abstract}
In the present work, the exact solution of Einstein's field equations which has been given by Curzon in 1924 \cite{Czn24} representing the field of a static binary system is reviewed. An adapted version of this solution is obtained to describe a dynamical binaries in a rotating coordinate system. It is shown that this version of the solution is time-dependent. It
reduces to the later one in the static case if the rotation goes to zero. The original Curzon solution shows that there are two
singularities at the two masses, while in the modified version the singularities become on the world-line of the two masses. The
solution shows no additional coordinate singularities. The killing vector field of the axial symmetry is obtained in the modified
version. In addition, the rotation admits a further rotational symmetry, so a rotation killing vector field is also obtained and
discussed. The equations of motion for a test particle in the field of a binary system are formulated and solved. Such equations have been used to study the gravitational time delay of arrival (Shapiro delay) of signals from binary pulsar systems resulted from our suggested modifications containing additional terms. These terms are interpreted as higher order corrections to the masses. In particular we investigate the gravito-magnetic effect due to orbital angular motion of the double pulsars. We give numerical estimates of this type of the time delay in the case of the double-pulsar system PSR J0737-3039 A\&B. We draw a model curve for the gravito-magnetic time delay during one orbital revolution. We suggest that this type of delay will have a larger contribution during the last phase of the system evolution.
\end{abstract}
\section*{Introduction}
What is meant by the two body problem is the problem of two structurless non-spinning point-like particles, characterized by two
mass parameters $m_{1}$ and $m_{2}$, moving under their mutual gravitational interaction. In order to formulate this problem, one
aims to get an explicit expression of acceleration of the binaries in terms of their positions and velocities. To discuss this problem we can dissolve it into two aspects: i) obtaining the equations of motion of the two interacting bodies, ii) solving these equations. In the context of Newtonian gravity the members of the binary system are considered as widely separated objects, such that the contribution of the non-linear effects can be neglected. Newtonian gravity has a linear structure that enables us to derive the equations of motion of the binary components, and also to get an exact solution for these equations. It gives a full treatment of the binary system. On the other hand, in General Relativity, the field equations have a non-linear hyperbolic structure, so that it is not easy even to get the equations of motion for such systems. Since in General Relativity the equations of motion are embedded in the field equations, consequently it is very difficult to derive the field equations as a linear functional of the matter distribution independently of the equations of motion \cite{LB01, DD81}.\\

One may ask why we are obliged to deal with General Relativity. Actually the Newtonian theory is very limited, since it deals only with systems, have a widely separated and slowly moving objects. Therefore, the later can not predict the behavior of systems with a small separation and fast moving objects. Moreover it can not account for the accumulating small feed back effects, e.g. the
advance of perihelia of planetary orbits. Fortunately, in such cases we are saved by General Relativity, which gives a good agreement with the observational and experimental tests. In order to avoid the above mentioned difficulties, most of the relativistic treatments in the literature have used an approximation technique, e.g. the \textit{Parameterized Post Newtonian} (PPN) method. This method expands the relativistic effects in power series of $v/c$, when only the relativistic expressions with order of $v^{2}/c^{2}$ added to the Newton's law, so it is called first post-Newtonian (1 PN) approximation \cite{Will93}. Recently, the equations of motion of binary systems have been derived at the third and half post-Newtonian (3.5 PN) order. However the beauty of this technique is its applicability for various classes of relativistic theories of gravity and also on its attractive mathematical formalism. The weak point of this technique is the violation of the general covariance principle.\\

Exact solutions in General Relativity have played an important role
to give a full description of some physical problems (e.g.
Schwarzschild, kerr, Reissner-Nordstr\"{o}m). The meaning of an
$``$\textit{Exact Solution}$"$ is that the metric is written in a coordinate system, in terms of the well-known analytical
functions. Unfortunately there are few exact solutions, for real
physical problems, that have been established, as mentioned by
Kinnersky c.f. \cite{KSMH80} $``$\textit{Most of the known exact
solutions describe situations which are frankly unphysical}$"$.\\
Although we have a lot of these non-physical solutions, one of the
most important problems, which remained without an exact solution,
is the physical field of two bodies. In general, for any physical
problem we aim to construct a mathematical model; using some
reasonable conditions; defined by a certain set of differential
equations. Actually it is not easy to interpret the obtained
solution, in some physical theories such as General Relativity,
because of its high non-linearity. But even if we could not
understand the qualitative features of an exact solution, we can
compare it with approximate results, which will be very useful, to
check the validity of this solution.\\

In the literature there are many attempts to solve the two body problem in the frame work of General Relativity. The solution required is expected to represent the field of two revolving objects (singularities) in one system. Among these attempts is that given by Curzon \cite{Czn24}. Unfortunately, this solution is static. \\

The aim of the present work is to adapt Curzon exact solution, of Einstein's field equations, to represent a time-dependent case of a binary system. This is done by rewriting the Curzon metric in a rotation frame. In general, rotation is rather poorly treated in general relativity, although it does represent a big conceptional problem which is called "Anti-Machian" nature \cite{Bondi97}. The rotating frame effect is essentially reduced to affecting the dynamical mass of the rotating body, exactly as the translational motion does, and to the spacetime geometry through gravito-magnetic effects. In Section \ref{Rotation and Space-time} we discuss the important role of rotating frames of reference in the context of general relativity. Also to show that the difference between rotating systems (non-inertial) and inertial systems does not only lie in a different category of the coordinates, but rather in different global chronogeometric properties of the choice of the reference frames. However the rotating systems do not identify any class of equivalent observers, which serves as an (absolute) reference frame. In 1924 Curzon obtained a solution, to Einstein's field equations, representing the field of two static singularities. For this reason, we give a brief review of Curzon solution in Section \ref{curzon Solution}, with its necessary background.\\

Unfortunately, this solution has been classified among un-physical solutions, since it does not represent a real astrophysical system. We are going to adjust this solution in order to be suitable for describing a binary system. This certifies a difficult task, but we are going to simplify this problem in the following manner. We assumed that the two members of the system have comparable masses rotate in a circular orbit about the center of mass of the system. We assumed further, that the line of sight is in the plane of the orbit. This represents a two body problem which represents many physical configurations especially the case of PSR J0737-3039 \cite{Bur03, Lyne04}. In Section \ref{A Modification of Curzon Solution}, we are going to adjust this solution to represent the field of a real binary system.\\

We test the solution by reducing it to a flat space and schwarzschild solutions, under certain conditions, in Section \ref{Check of the new solution}, also the coordinate singularities and the additional rotational killing vector field admitted has been obtained and discussed.\\

The motion, of a test particle, in the modified Curzon field is treated in Section \ref{Motion in The Modified Curzon Field}. The equations of motion are applied to find the time delay of the pulses of a pulsar in a binary system and discussed in the case of the double pulsar PSR J0737-3039 like, in Section \ref{Time delay in binary systems}. The work is discussed and concluded in section \ref{Prospective}.
\section{Rotation and Space-time}\label{Rotation and Space-time}
Some authors claim that the rotating reference frame is merely a coordinate transformation. In this section we are going to show that the new possible description of the physical processes in the rotating system is not fake consequences of the coordinate transformation. It is the space-time geometry itself that is at the issue. It seems that there is a common agreement on the special theory of relativity and its conceptual foundations and experimental results. But one of the important points is the effect of a rotating frame of reference which is still misunderstood. In fact, in 1909 Ehrenfest pointed out to an internal contradiction in the SR applied to the case of a rotating disk. Few years later, Sagnac in 1913, showed a contradiction of special relativity with experimental data.\\

The peculiarities of rotating frames have led Einstein to stop his trails to generalize the Special theory of relativity by considering accelerating frames. The consideration of coordinates on a rotating disc played the leading role in the genesis of general relativity. It turns out, the difference between \textit{inertial} and \textit{non-inertial} frames of reference, and between the special and the general relativity, is not in the choice of the coordinates. Rather, the difference is that $``$\textit{chronogeometric characteristics become globally different}$"$ \cite{D10}.
\subsection{The Sagnac Effect Experiment}
Now assume that two signals of light are emitted from a  fixed source in a rotating frame and begin traveling, in opposite directions, along the same ring trajectory with a constant radius $r$. We follow the two signals locally using a successive local co-moving frames, the elapsed time can be calculated by integrating the time intervals as measured by the these frames in which one can hold the constant speed of light $c$. We may expect that the two signals use the same time interval to complete their circular trajectories and return back to their source. But, experiment shows that the arrival times of the two signals are different. However, because of the time gaps $\Delta t$ the two signals do not complete their circles simultaneously.\\

As seen above the Sagnac effect does not affected by a specific nature of the signals that propagate in the two directions. But, it reflects directly the space-time geometry of the rotating reference frame. So we conclude that the same Sagnac time difference is there for any two identical signals traveling into two directions as well as light. The Sagnac experiment directly investigates the the relation between space-time relations and the rotating reference frame.\\

This led some authors to attempt a non-time-orthogonal analysis \cite{K03}. They have concluded that not only the sagnac effect be derived, but also all other observed rotating frame effects can be derived as well. Others such experts as Neil Ashby goes further to violate constancy of the speed of light in a rotating frame, and holding the speed of light as a constant in rotating frames leads to a significant error (c.f. \cite{K03} and the references therein [15-25]).
\subsection{Measuring Concept in 3+1 Space-time}
 The second postulate of the special relativity states that "laws of Nature remain invariant relative to all inertial reference frames". On the other hand any "observable" physical quantity is in general frame-dependent! Then a problem arises. To over come this fake conflict one can say the physical quantities may vary from frame to another, but they would always combine in a such way to keep the physical law the same for all observers. In the mathematical model of General Relativity, physical quantities are expressed by world tensors, to grantee the covariance principle for all physical laws, which are just relations among these quantities. So, given a reference frame, or how do we relate these absolute quantities to the relative ones? And how do we relate physical laws to reference-dependent ones? Actually this can be done by a suitable 3+1 splitting, the mathematical model of spacetime to the observable quantities which are relative to a reference frame?\\

This led some authors \cite{RR02} to discuss the process of the a splitting procedure needed to obtain quantities that have a true physical meaning, i.e. which are gauge invariant and, hence, observable. This has been done by using the Cattaneo's projection technique to study the curvature invariance in the 'relative space' of a rotating platform. Then they have show that the geometry of the relative space of the disk has a space curvature which is not zero\footnote{The vanishing of the curvature scalar of the spacetime still the same for all observers according to the covariance principle. But the spatial curvature scalar has non-zero value in a rotating frame, which implies that a non-zero temporal scalar curvature affecting our measurements of time.}. We can summarize this in the following diagram:\\

\begin{centering}
$$\xymatrix{\mathop{\textmd{Space-time}}\limits_{4D}&R_{\alpha\beta\gamma\delta}=0\ar[d]|{projection}\ar[rr]^{rotation}&&R_{\alpha\beta\gamma\delta}=0\ar[rr]^{contraction}\ar[d]|{projection}&&R=0\\
          \mathop{\textmd{Spatial-space}}\limits_{3D}&R_{abcd}=0                                                     &&R_{abcd}\neq 0\ar[rr]^{contraction}            &&R \neq 0.}$$
\end{centering}

One of the most qualitative results of this technique is the indication of the existence of real physical effects depending only on the rotating frame. As it appears only when $\omega \neq 0$. It should be noted that the peculiarity of the description of physical processes in the rotating frame is not a fake consequences of the coordinate transformation \cite{K03, C07}. But The space-time geometry itself is at the issue.\\

Another interesting point when studying the invariants of the curvature tensor, i.e. the Kretschmann scalar $R_{abcd} \, {\! R}^{abcd}$ and the Chern-Pontryagin scalar\footnote{where ${{}^\star R}^{abcd}$ is the left dual of the Riemann tensor, similar to the electromagnetic field tensor $F_{ab} \, F^{ab}, \; \; F_{ab} \, {{}^\star \! F}^{ab}$.} $R_{abcd} \, {{}^\star \! R}^{abcd}$, we find a non-vanishing Kretschmann scalar but the Chern-Pontryagin scalar vanishes for the static Schwarzschild metric at rest frame. Also for the static metric boosted by velocity $v$ the  Chern-Pontryagin scalar scalar is still zero. While, considering a steady rotating source gives non-vanishing values for both curvature invariants \cite{JF04}. Now we may summarize this section in few points:
\begin{enumerate}
\item [(1)] The difference between translation (\textit{i.e. inertial}) and rotation (\textit{i.e. non-inertial}) frames is physical not a fake consequences of coordinate transformation.
\item [(2)] The rotating frames affecting the global structure of the space-time.
\item [(3)] Some scalar invariants, in the four dimensional space-time, may be changed due to consideration of a rotating frame of reference. This indicates different space-time geometry.
\item [(4)] The natural split from four dimensional space-time to 3+1 leads to a non-vanishing spatial scalar curvature, which implies a non-vanishing temporal scalar curvature affecting our measurements of time.
\end{enumerate}
This showed that the rotating frames of reference may play an important role to interpret (modify) some un-physical solutions of the general theory of relativity. Among this un-physical solutions the Curzon metric \cite{Czn24}, which is suggested originally to describe the general relativistic gravitational field of two objects.
\section{Curzon Solution}\label{curzon Solution}
This solution is a Weyl class one, which has been found soon after the
birth of GR. It refers to two singularities on the axis of
symmetry. This is mentioned by Bonnor \cite{Bon92} as $``$Probably
the most perspicacious of all exact solutions in GR$"$.
\subsection{Standard Curzon Static Field}\label{Standard Curzon Static Field}
In this section, we aim to review the solution, obtained by Curzon
in 1924 \cite{Czn24}, of Einstein's equations
\begin{equation}\label{empty field eqn}
\bar{R}_{\alpha \beta} =0,
\end{equation}
for an axial symmetric gravitational field produced by two
singularities ($A$, $B$) on the axis $\bar{x}^{1}$ separated by a distance
$2a$. If the origin of a reference frame lies at the mid point
between the two singularities on this axis, so for an arbitrary
point $P$ in the plane of $\bar{x}^{1}~ \textmd{and}~ \bar{x}^{2}$
coordinates the direction of these singularities can be defined
using bipolar coordinates $\bar{r}_{1}~ \textmd{and}~ \bar{r}_{2}$. Where
\begin{equation}\label{q2.2}
    \bar{r}_{1}^{2} =\left( \bar{x}^{1}-a\right) ^{2}+\left( \bar{x}^{2}\right) ^{2},~   \bar{r}_{2}^{2} =\left( \bar{x}^{1}+a\right) ^{2}+\left( \bar{x}^{2}\right) ^{2}.
\end{equation}
The cylindrical coordinates used in this study are
\begin{equation}\label{q2.3}
\bar{x}^{1}=\bar{z},~~~\bar{x}^{2}=\bar{\rho},~~~
\bar{x}^{3}=\bar{\phi},~~~ \bar{x}^{4}=\bar{t}.
\end{equation}
The metric characterizing the space, with axial-symmetric static
gravitational field \cite{Czn24}, is
\begin{equation}\label{q2.4}
d\bar{s}^{2}=-e^{\bar{\mu}}\left( d\bar{z}^{2}+d\bar{\rho
}^{2}\right) -e^{-\bar{\nu }}\bar{\rho }^{2}d\bar{\phi
}^{2}+e^{\bar{\nu }}d\bar{t}^{2},
\end{equation}
where $\bar{\mu}\equiv \bar{\mu}\left( \bar{z},\bar{\rho
}\right),~ \bar{\nu }\equiv \bar{\nu }\left( \bar{z},\bar{\rho }\right).$\\

Curzon has found the following solutions for the above form that represents a static
two-particle system, by setting
\begin{eqnarray}\label{q2.8}
\bar{\nu}&=&-2\left[\frac{m_{1}}{\bar{r}_{1}}+\frac{m_{2}}{\bar{r}_{2}}\right],\label{linfld}\\
\bar{\mu}&=&2\left[\frac{m_{1}}{\bar{r}_{1}}+\frac{m_{2}}{\bar{r}_{2}}\right]
-\left[\dfrac{m_{1}^{2}}{\bar{r}_{1}^{4}}+\dfrac{m_{2}^{2}}{\bar{r}_{2}^{4}}\right]
\bar{\rho}^{2}+\dfrac{m_{1}m_{2}}{a^{2}}\left[\left({\bar{z}^{2}+\bar{\rho}^{2}
-a^{2}}\over{\bar{r}_{1}\bar{r}_{2}}\right)-1\right].\label{quadfld}
\end{eqnarray}
where $m_{1}~\textmd{and}~m_{2}$ are constants of the integration.
One may realize how the contribution of the classical theory of
gravity is represented by this linear equation (\ref{linfld}), while the general relativistic effects of the curved space-time is given by the quadratic terms in equation (\ref{quadfld}).\\
Now by defining two angles $\bar{\alpha}_{1}$ and $\bar{\alpha}_{2}$, as the angles between by $AP$, $BP$ and the axis $\bar{z}$,
respectively, one can write the solution of (\ref{empty field eqn}), in the present case, as
\begin{eqnarray}
\bar{\nu}&=&-2\left[\dfrac{m_{1}}{\bar{r}_{1}}+\dfrac{m_{2}}{\bar{r}_{2}}\right],\label{q2.16}\\[10pt]
\bar{\mu}&=& 2\left[\dfrac{m_{1}}{\bar{r}_{1}}+\dfrac{m_{2}}{\bar{r}_{2}}\right]
            -\left[\dfrac{m_{1}^{2}}{\bar{r}_{1}^{2}}~sin^{2}{\bar{\alpha}_{1}}
            +\dfrac{m_{2}^{2}}{\bar{r}_{2}^{2}}~sin^{2}{\bar{\alpha}_{2}}\right]\notag\\[3pt]
         & &-2\dfrac{m_{1}m_{2}}{a^{2}}~sin^{2}\left(\tfrac{\bar{\alpha}_{1}-\bar{\alpha}_{2}}{2} \right).\label{q2.17}
\end{eqnarray}
The solution given by (\ref{q2.16}) and (\ref{q2.17}) is the solution given by Curzon in its standard form \cite{Czn24}.
\subsection{Structure of the Curzon Metric}
Many authors prefer \cite{RN68, S36, S85, SM73} to write the solution in an equivalent form for studying the structure of Curzon solution. This is done by introducing a new function $\bar{\lambda}$, such that
\begin{equation}\label{q2.9}
\bar{\lambda}=\bar{\mu}+\bar{\nu},
\end{equation}
Curzon solution can be rewritten in the following equivalent form.
\begin{equation*}\label{eq:czm}
    ds^{2}=e^{\bar{\nu}}d\bar{t}^{2}-e^{-\bar{\nu}}\left[e^{\bar{\lambda}}\left(d\bar{z}^{2}+d\bar{\rho}^{2}\right)
    +\bar{\rho}^{2}d\bar{\phi}^{2}\right],
\end{equation*}
where
\begin{eqnarray*}
\bar{\nu} &=&-2\left[\dfrac{m_{1}}{{\bar{r}}_{1}}+\dfrac{m_{2}}{{\bar{r}}_{2}}\right],\\[10pt]
\bar{\lambda}&=&-\left[\dfrac{m_{1}^{2}}{{\bar{r}}_{1}^{4}}+\dfrac{m_{2}^{2}}{{\bar{r}}_{2}^{4}}\right]{\bar{\rho}}^{2}
            +\dfrac{m_{1}m_{2}}{a^{2}}\left[\left({{\bar{z}}^{2}+{\bar{\rho}}^{2}-a^{2}}\over{{\bar{r}}_{1}{\bar{r}}_{2}}
            \right)-1\right].
\end{eqnarray*}
This solution was the source of a debate between Einstein and Silberstein \cite{S36, ER36}, the latter claiming
that the solution indicated the incorrectness of general relativity since its field equations yielded a
patently unphysical solution: two static point singularities completely surrounded by vacuum. The physical interpretation of the obtained solution, that the region $\bar{\rho} = 0,~\abs{\bar{z}} < a$ cannot be considered as a vacuum solution of the Einstein equations. From the current standpoint of general relativity the only alternative is to postulate a $T_{\mu\nu} \neq 0$ within this region, i.e. to introduce a \textit{strut} \cite{S85}.\\

To compute the force between the two singularities, where $\bar{\lambda}(0) \neq 0$ along the $z$-axis between the two
singularities, can be given as
\begin{equation}\label{q2.18}
\bar{\lambda}=\dfrac{m_{1}m_{2}}{a^{2}}\left[\dfrac{\bar{z}^{2}-a^{2}}{\bar{r}_{1}\bar{r}_{2}}-1\right]=
-2\dfrac{m_{1}m_{2}}{a^{2}},
\end{equation}
then the stress force between the two masses
\begin{equation}\label{q2.19}
F=-\dfrac{GM_{1}M_{2}}{(2d)^{2}},
\end{equation}
if we take the Newtonian limit $d\gg m_{1},m_{2}$, the medium between the two masses contains a compression merely the Newtonian force attraction \cite{RN68}. This indicating (or suggesting) necessarily existence of singular structures ("struts", "ropes", or "membranes") that are responsible for holding masses against the attractive force of gravity in a static configuration. This gives impression that the field equations in general relativity involve also equations of motion.\\

Einstein pointed out that the line element represents a regular gravitational field outside of the two particles when $\bar{\nu}$ and $\bar{\lambda}$ and their first derivatives are continuous and also $\bar{\lambda}$ must vanish everywhere for $\bar{\rho}=0$ except at the two mass-points. But the nonvanishing value of $\bar{\lambda}$ on the axis ($\bar{\rho} =0$) between the two mass-points does not satisfy the regularity conditions. This led Einstein to say that the above solution must be ruled out as a purely vacuum solution because of the following consideration. Consider a circle with center at $\bar{\rho} = 0$ in the two-dimensional subspace $\bar{t} = const,~\bar{z} = const$ with $\abs{\bar{z}} < a$. If we take the limit of the ratio of its circumference $C$ to its diameter $D$ as $D \To 0$, we find that $C/D \To \pi e^{-\bar{\lambda}(0)}$. Since $\bar{\lambda}(0) \neq 0$ for $\abs{\bar{z}} < a$, $C/D$ does not approach $\pi$, and hence the above spacetime violates the condition of elementary flatness \cite{ER36, S85}.\\

Also one of the interesting points is the directional singularity at $\bar{\rho}^{2}+\bar{z}^{2}=0$. For example, the limit of the Kretschmann scalar invariant $\bar{R}^{\alpha\beta\gamma\delta}\bar{R}_{\alpha\beta\gamma\delta}$ depends on the direction of approach to singularity. Then the behaviour of invariants of the curvature tensor in the limit $\bar{R} \To 0$ is strongly dependent on the direction of approach. In other words, the limit of such invariants along the $\bar{z}$-axis is in fact regular, while it is singular along other directions; which leads one to make the suggestion that possibly the Curzon metric "opens up" for particles approaching $\bar{R} = 0$ along the $\bar{z}$-axis, allowing them to pass on into some new region \cite{SM73}.

Another remarkable point for this solution is showing that a different behavior with respect to the manifold from the Schwarzchild metric. In the case of negative mass both show point-like singularity \cite{S68}. In the case of positive mass the Schwarzschild metric can have a nonsingular event horizon, while the Curzon metric shows a singular event horizon of infinite area.This shows a distinct topological background for positive and negative mass in the two metrics. Although the Curzon metric has axial symmetry, the Schwarzschild solution (a spherically symmetric metric) can be found as a special case!\\

Moreover, the solution is an exact solution for which one can find such quantities of physical interest as radiation patterns. Also it is not necessary to require the weak-field initial data. For all these characteristics, one may find some interest in this solution.
\section{Modification of Curzon Solution}\label{A Modification of Curzon Solution}
Actually the static case, which had been studied by Curzon, does not represent any physical configuration (e.g. real binary systems in Nature) in view of the fact that the binary systems are always dynamical systems. The main aim of the present section is to show how to modify Curzon solution such that it can be used to represent the field of a real physical binary system. Many authors aimed to use techniques for generating new stationary solutions from the static ones \cite{HCD84, GS03}, including the Curzon solution, in order to give a physical interpretation for the Curzon static configuration. This is by considering the stationary system of two masses kept apart by their gravitational spin-spin interaction to stabilize the two masses by addition of angular momentum \cite{H90, DH82, DH85}. Actually these developed solutions do not represent any physical configuration in Nature.\\

In order to describe such systems, the effect of time should be introduced, by considering a co-rotating coordinate system,
associated with the angular motion of the collinear singularities, with respect to a frame of reference fixed at the center. It is similar to the situation of using co-moving coordinate systems in cosmological applications. Some claim that the rotating coordinates will not provide a new physics. This is right in the sense of the spacetime is absolute with respect to all observers, but all our measurements are carried out in space and in time. The spacetime is absolute with respect to all inertial observers, while the case of non-inertial observers is different as discussed in section \ref{Rotation and Space-time}. Usually, practical arguments require a specific choice of one coordinate system over another which may lead to different choices according to the situation. When geometrical relations become different, coordinate systems with different characteristics, adjusted to the new geometry, may lead to a simpler description. \emph{Although rotational motion has a character on its own since it appears to be absolute, unlike translational motion, which is purely relative}. But this does not change the conventional view of the coordinates' nature. The transition from inertial to non-inertial systems, or from special to general relativity, changes the global characteristics of the physical temporal and spatial relations.
\subsection{General Outlines of the Modification}\label{General Outlines of The Modification}
Curzon has chosen the point $P$ in the ($\bar{z}$-$\bar{\rho}$)
plane. Since $A$ and $B$ are static they could not define a
particular plane, he always can reorient the coordinates by choosing
a particular value of an angle $\bar{\phi}$ to make $ABP$ plane
always coincides with the ($\bar{z},\bar{\rho}$) plane, without
affecting generality. While the non-static case, such collinear
singularities $A$ and $B$ are rotating around a common center $C$ by
an angular velocity $\omega$. It is clear that the moving
singularities are defining a particular plane (orbital plane), so
the point $P$ should be an arbitrary point, not necessary in this
plane. The separations between this point and the two singularities
($A,B$) are respectively
\begin{equation}\label{q3.1}
\left.
\begin{split}
\bar{r}_{1}^{2}=\left(\bar{z}-a\right)^{2}+\bar{\rho}^{2}\sin^{2}\bar{\phi}, \\
\bar{r}_{2}^{2}=\left(\bar{z}+a\right)^{2}+\bar{\rho}^{2}\sin^{2}\bar{\phi}.
\end{split}
\right\}
\end{equation}
Where the angle $\bar{\phi}$ represents the inclination of the
orbital plane ($\bar{z}, \bar{\rho}$); in the case of a rotating system; to the plane of the sky. As $\bar{\phi} \rightarrow
\pi/2$ the binary system tends to be an eclipsing binary system with respect to an observer at the point $P$.\\

In general the singularities (binary components) are assumed to be
separated from $C$ by distances $a$ and $b$ ($a \not= b$), and the
angular velocity $\omega$ in this case is a function of time. To
simplify the problem we are going to assume that the motion is
circular as following:
\begin{enumerate}
    \item The two singularities have comparable masses.
    \item The distances, $a$ and $b$, are equal.
    \item The angular velocity $\omega$ of the collinear
    singularities is constant.
    \item The orbits of the two singularities will coincident on
    each other producing a circular orbit with a radius $a$.
\end{enumerate}
Now before discussing the non-static case, we will return to
standard form of Curzon metric (\ref{q2.4}), and show its form in
other coordinate systems. This is done in order to facilitate
comparison with some special cases.
\subsection{A Particular Choice of Coordinate Systems}\label{A Particular Choice of a Coordinate System}
In order to refer the points of the manifold to the Cartesian
coordinate system, as general covariance is still preserved since we
use tensors in the formalism, we use the transformation
$$TI \colon (\bar{z},\bar{\rho},\bar{\phi},\bar{t}) \To (\tilde{x},\tilde{y},\tilde{z},\tilde{t})$$
i.e.
\begin{equation}\label{q3.2} \left.
\begin{split}
\bar{z} &= \tilde{z}, \\
\bar{\rho} &= \sqrt{\tilde{x}^{2}+\tilde{y}^{2}},\\
\bar{\phi} &= \tan^{-1}(\tilde{x}/\tilde{y}), \\
\bar{t} &= \tilde{t}.
\end{split}
\right\}
\end{equation}
Applying (\ref{q3.2}) to Curzon metric (\ref{q2.4}) and recalling
that $ds$ is a scalar, we get
$$d\bar{s}^{2}(\bar{x}^{\beta})=d\tilde{s}^{2}(\tilde{x}^{\alpha})$$
So we can rewrite (\ref{q2.4}) in the form
\begin{eqnarray}
d\tilde{s}^{2}&=& -\dfrac{\tilde{x}^{2}e^{\tilde{\mu}}+\tilde{y}^{2}}{\tilde{x}^{2}
+\tilde{y}^{2}}~d{\tilde{x}^{2}}-\dfrac{2\tilde{x}\tilde{y}}{\tilde{x}^{2}+\tilde{y}^{2}}
(e^{\tilde{\mu}}-e^{\tilde{\nu}})~d{\tilde{x}}~d{\tilde{y}}\notag\\[3pt]
              & & -\dfrac{\tilde{y}^{2}e^{\tilde{\mu}}+\tilde{x}^{2}e^{-\tilde{\nu}}}
              {\tilde{x}^{2}+\tilde{y}^{2}}~d{\tilde{y}^{2}}-e^{\tilde{\mu}}~
              d{\tilde{z}^{2}}+e^{\tilde{\nu}}~d{\tilde{t}^{2}},\label{q3.3}
\end{eqnarray}
where
\begin{subequations}
\renewcommand{\theequation}{\theparentequation.\arabic{equation}}
\begin{eqnarray}
\tilde{\nu} \left( \tilde{x},\tilde{y}, \tilde{z}\right)&=&
-2\left[\dfrac{m_{1}}{\tilde{r}_{1}}+\dfrac{m_{2}}{\tilde{r}_{2}}\right]\label{q3.4.1}\\[10pt]
\tilde{\mu}\left(\tilde{x},\tilde{y},\tilde{z}\right)&=&
2\left[\dfrac{m_{1}}{\tilde{r}_{1}}+\dfrac{m_{2}}{\tilde{r}_{2}}\right]
-\left(\tilde{x}^{2}+\tilde{y}\right)\left[\dfrac{m_{1}}{\tilde{r}_{1}^{4}}+\dfrac{m_{2}}
{\tilde{r}_{2}^{4}}\right]\notag\\[3pt]
                & &+\dfrac{m_{1}m_{2}}{a^{2}}\left[\left(\dfrac{\tilde{x}^{2}
                +\tilde{y}^{2}+\tilde{z}^{2}-a^{2}}{\tilde{r}_{1}\tilde{r}_{2}}\right)-1\right],\label{q3.4.2}
\end{eqnarray}
\end{subequations}
and
\begin{equation}\label{q3.5}
\left.
\begin{split}
\tilde{r}_{1}^{2}& = & \tilde{x}^{2}+\tilde{y}^{2}+\tilde{z}^{2}+a^{2}-2a\tilde{z}\\
\tilde{r}_{2}^{2}& = &
\tilde{x}^{2}+\tilde{y}^{2}+\tilde{z}^{2}+a^{2}+2a\tilde{z}.
\end{split}
\right\}
\end{equation}
The metric (\ref{q3.3}) represents the gravitational field of two
singularities, and has axial symmetry about the $\tilde{z}$-axis.\\

We next write the solution in the spherical polar coordinates. This may be not adapted to the symmetry of the solution in the static case but it will be beneficial in the no-static case. Now we are going to apply a second transformation to represent the
metric (\ref{q3.3}) in spherical polar coordinate, as follows
$$TII \colon (\tilde{x},\tilde{y},\tilde{z},\tilde{t}) \To (\hat{r},\hat{\theta},\hat{\phi},\hat{t})$$
i.e.
\begin{equation}\label{q3.6}
\left.
\begin{split}
\tilde{x} & = \hat{r} ~\cos\hat{\theta}\\
\tilde{y} & = \hat{r} ~\sin\hat{\theta} ~\cos\hat{\phi}\\
\tilde{z} & = \hat{r} ~\sin\hat{\theta} ~\sin\hat{\phi}\\
\tilde{t} & = \hat{t}
\end{split}
\right\}
\end{equation}
where $0 \leq \theta \leq \pi$ and $0 \leq \phi \leq 2\pi$. One can
notice that the transformation $TII$ is not the conventional
transformation. Consequently, the angle $\hat{\phi}$ is not the same
angle $\bar{\phi}$, given in the standard Curzon
solution.\\
Since,
$$d\hat{s}^{2}(\hat{x}^{\gamma})=d\tilde{s}^{2}(\tilde{x}^{\alpha}).$$
Then, the metric coefficient will be given by
\begin{subequations}
\renewcommand{\theequation}{\theparentequation.\arabic{equation}}
\begin{eqnarray}
g_{\hat{r}\hat{r}}\vspace{2cm}&=&-e^{\hat{\mu}}\label{q3.7.1}\\[10pt]
g_{\hat{\theta} \hat{\theta}}\vspace{2cm}&=&-\hat{r}^{2}~\dfrac{\cos^{2}\hat{\theta}~\sin^{2}\hat{\phi}~
e^{\hat{\mu}}+\cos^{2}\hat{\phi}~e^{-\hat{\nu}}}{\cos^{2}{\hat{\theta}}+\sin^{2}\hat{\theta}
~\cos^{2}\hat{\phi}}\label{q3.7.2}\\[10pt]
g_{\hat{\theta} \hat{\phi}}\vspace{2cm}&=&-\dfrac{~\hat{r}^{2}\sin\hat{\theta}~\sin\hat{\phi}~\cos\hat{\theta}~
\cos\hat{\phi}}{\cos^{2}{\hat{\theta}}+\sin^{2}\hat{\theta}
~\cos^{2}\hat{\phi}}\left(
e^{\hat{\mu}}-e^{-\hat{\nu}}\right)\label{q3.7.3}\\[10pt]
g_{\hat{\phi} \hat{\phi}}\vspace{2cm}&=&-\hat{r}^{2}~\sin^{2}\hat{\theta}~\dfrac{\cos^{2}\hat{\theta}~
\sin^{2}\hat{\phi}~e^{-\hat{\nu}}+\cos^{2}\hat{\phi}~e^{\hat{\mu}}}{\cos^{2}{\hat{\theta}}+\sin^{2}\hat{\theta}
~\cos^{2}\hat{\phi}}\label{q3.7.4}\\[10pt]
g_{\hat{t}\hat{t}}&=&e^{\hat{\nu}}\label{q3.7.5}
\end{eqnarray}
\end{subequations}
where
\begin{subequations}\label{q3.8}
\renewcommand{\theequation}{\theparentequation.\arabic{equation}}
\begin{eqnarray}
\hat{\nu} ( \hat{r},\hat{\theta},\hat{\phi} )&=&
-2\left[\dfrac{m_{1}}{\hat{r}_{1}}+\dfrac{m_{2}}{\hat{r}_{2}}\right]\label{q3.8.1}\\[10pt]
\hat{\mu} ( \hat{r},\hat{\theta},\hat{\phi} )&=&
2\left[\dfrac{m_{1}}{\hat{r}_{1}}+\dfrac{m_{2}}{\hat{r}_{2}}\right]\notag\\[3pt]
&&-\hat{r}^{2}\left(\cos^{2}\hat{\theta}+\sin^{2}\hat{\theta}~\cos^{2}
\hat{\phi}\right)\left[\dfrac{m_{1}}{\hat{r}_{1}^{4}}+\dfrac{m_{2}}{\hat{r}_{2}^{4}}\right]\notag\\[3pt]
                                                                 & &
+\dfrac{m_{1}m_{2}}{a^{2}}\left[\left(\dfrac{\hat{r}^{2}-a^{2}}{\hat{r}_{1}\hat{r}_{2}}\right)-1\right]\label{q3.8.2}
\end{eqnarray}
\end{subequations}
and
\begin{equation}\label{q3.9}
\left.
\begin{split}
\hat{r}_{1}^{2}&=(\hat{r}+a)^{2}-2a\hat{r}(1+\sin\hat{\theta}~\sin\hat{\phi})\\
\hat{r}_{2}^{2}&=(\hat{r}+a)^{2}-2a\hat{r}(1-\sin\hat{\theta}~\sin\hat{\phi})
\end{split}
\right\}
\end{equation}
The solution in this form is ready to be written in a rotating reference frame. In view of the previous illustration, considering the non-static case assumptions, given in $\S$\ref{General Outlines of The Modification}, we apply the third transformation
$$TIII \colon (\hat{r},\hat{\theta},\hat{\phi},\hat{t}) \To (r,\theta,\phi,t)$$
The following rotating platform defines a non-time-orthogonal physical frame (because of the time-dependence of $phi$), unlike the stationary (i.e. inertial) case; i.e.
\begin{equation}\label{q3.10}
\left.
\begin{split}
\hat{r} & = r,\\
\hat{\theta} & = \theta,\\
\hat{\phi} & = \phi+\omega t,\\
\hat{t} & = t,
\end{split}
\right\}
\end{equation}
which if combined with the transformation law of the scalar,
$$ds^{2}(x^{\sigma})=d\hat{s}^{2}(\hat{x}^{\gamma}),$$
would give the following non-vanishing components of the metric
tensor in terms of the new coordinate system ($r$, $\theta$, $\phi$,
$t$).
\begin{subequations}\label{q3.11}
\renewcommand{\theequation}{\theparentequation.\arabic{equation}}
\begin{eqnarray}
g_{rr}&=&-e^{\mu},\label{q3.11.1}\\[10pt]
g_{\theta \theta}&=&-r^{2}~\dfrac{\cos^{2}\theta~\sin^{2}(\phi+\omega
t)~e^{\mu}+\cos^{2}(\phi+\omega
t)~e^{-\nu}}{\left[\cos^{2}{\theta}+\sin^{2}{\theta}~\cos^{2}(\phi+\omega
t)\right]},\label{q3.11.2}\\[10pt]
g_{\theta \phi}&=&-\dfrac{~r^{2}\sin{\theta}~\sin{(\phi+\omega
t)}~\cos{\theta}~\cos{(\phi+\omega
t)}}{\left[\cos^{2}{\theta}+\sin^{2}{\theta}~\cos^{2}(\phi+\omega
t)\right]}\left(
e^{\mu}-e^{-\nu}\right),\label{q3.11.3}\\[10pt]
g_{\theta t}&=&-\omega \dfrac{~r^{2}\sin{\theta}~\sin{(\phi+\omega
t)}~\cos{\theta}~\cos{(\phi+\omega
t)}}{\left[\cos^{2}{\theta}+\sin^{2}{\theta}~\cos^{2}(\phi+\omega
t)\right]}\left(
e^{\mu}-e^{-\nu}\right),\label{q3.11.4}\\[10pt]
g_{\phi \phi}&=&-r^{2}
\sin^{2}{\theta}~\dfrac{\cos^{2}\theta~\sin^{2}(\phi+\omega
t)~e^{-\nu}+\cos^{2}(\phi+\omega
t)~e^{\mu}}{\left[\cos^{2}{\theta}+\sin^{2}{\theta}~\cos^{2}(\phi+\omega
t)\right]},\label{q3.11.5}\\[10pt]
g_{\phi t}&=&-\omega r^{2}
\sin^{2}{\theta}~\dfrac{\cos^{2}\theta~\sin^{2}(\phi+\omega
t)~e^{-\nu}+\cos^{2}(\phi+\omega
t)~e^{\mu}}{\left[\cos^{2}{\theta}+\sin^{2}{\theta}~\cos^{2}(\phi+\omega
t)\right]},\label{q3.11.6}\\[10pt]
g_{tt}&=&-{\left[\omega^{2} r^{2}
         \sin^{2}{\theta}\left(\cos^{2}\theta~\sin^{2}(\phi+\omega
         t)~e^{-\nu}+\cos^{2}(\phi+\omega
         t)~e^{\mu}\right)\right.}\notag\\[10pt]
      & &-{\left.\left(\cos^{2}{\theta}+\sin^{2}{\theta}~\cos^{2}(\phi+\omega
         t)\right)e^{\nu}\right]}/\notag\\[10pt]
      & &{\left[\cos^{2}{\theta}+\sin^{2}{\theta}~\cos^{2}(\phi+\omega
         t)\right]},\label{q3.11.7}
\end{eqnarray}
\end{subequations}
while (\ref{q3.8.1}), (\ref{q3.8.2}) and (\ref{q3.9}) now read:
\begin{subequations}\label{3.12}
\renewcommand{\theequation}{\theparentequation.\arabic{equation}}
\begin{eqnarray}
\nu \left( r,\theta ,\phi ,t \right)&=&
-2\left[ \dfrac{m_{1}}{r_{1}}+\dfrac{m_{2}}{r_{2}}\right],\label{q3.12.1}\\[10pt]
\mu \left( r,\theta ,\phi ,t \right)&=&
2\left[\dfrac{m_{1}}{r_{1}}+\dfrac{m_{2}}{r_{2}}\right]\notag\\
                                               & &
-{r}^{2}\left(\cos^{2}{\theta}+\sin^{2}{\theta}~\cos^{2}{(\phi+\omega t)}\right)
\left[ \dfrac{m_{1}^{2}}{r_{1}^{4}}+\dfrac{m_{2}^{2}}{r_{2}^{4}}\right] \notag\\
                                               & &
+\dfrac{m_{1}m_{2}}{a^{2}}\left[
\left(\dfrac{r^{2}-a^{2}}{r_{1}r_{2}}\right)-1\right],\label{q3.12.2}
\end{eqnarray}
\end{subequations}
and
\begin{equation}\label{q3.13}
\left.
\begin{split}
{r}_{1}^{2}&=({r}+a)^{2}-2a{r}(1+\sin{\theta}~\sin{(\phi+\omega t)}),\\
{r}_{2}^{2}&=({r}+a)^{2}-2a{r}(1-\sin{\theta}~\sin{(\phi+\omega
t)}).
\end{split}
\right\}
\end{equation}
Thus the set of quantities (\ref{q3.11.1})-(\ref{q3.11.7}) represents the field which is produced by a dynamical two-body
system, at any chosen point outside the two singularities. \emph{We have to mention here the introduced rotational motion provide a completely different view for the binary system. However the treatment of the translational motion incorporates the very notion of inertial reference frames and inertial observers, whereas rotating systems do not identify any class of equivalent observers. Also from the general relativity view we know that a mass curves space time around. If the source of the gravitational field rotates, the peculiar motion introduces further warps in spacetime, producing expectedly measurable effects on spacetime. This is important because it could give an opportunity to verify a general relativistic effect caused by the angular momentum of a source of gravitational field.}
\section{Boundary Conditions on the Field}\label{Check of the new solution}
The line element of the modified Curzon solution is given in terms of spherical polar coordinates ($r$,$\theta$,$\phi$,$t$). In order to check this solution one should obtain the flat space metric and Schwarzschild metric as limiting cases under certain conditions.
\subsection{First Check: Flat Space Metric}\label{The First Check: Flat Space Metric}
It can be easily shown that the $\mu$ and $\nu$ functions, given by (\ref{q3.12.1}) and (\ref{q3.12.2}), vanish as $r_{1}$, $r_{2} \to \infty$, i.e. at a large distance from the binary system. In this case the metric coefficients (\ref{q3.11.1})-(\ref{q3.11.7}) will become
$$g_{rr}=-1,~g_{\theta \theta}=-r^{2},~g_{\phi \phi}=-r^{2}\sin^{2}{\theta},~g_{tt}=1$$
and the line element will reduce to
$$ds^{2}=-dr^{2}-r^{2}d\theta^{2}-r^{2}\sin^{2}{\theta}d\phi^{2}+dt^{2}$$
which describes the space in absence of the gravitational field.
\subsection{Second check: Schwarzschild Metric}\label{The Second Check: Schwarzschild Metric}
Taking $\omega=0$, $m_{1}+m_{2}=m$, and taking {r} large enough, $r_{1} \approx r_{2} \approx
r$, such that $\mathcal{O}\left(\dfrac{1}{r^{2}}\right) \to 0$. In
this case the metric coefficients (\ref{q3.11.1})-(\ref{q3.11.7})
will become
$$g_{rr}=-e^{\mu},~g_{\theta \theta}=-r^{2} e^{\mu},~g_{\phi \phi}=-r^{2} \sin^{2}{\theta} e^{\mu},~g_{tt}=e^{\nu},$$
and the line element can be written as
$$ds^{2}=-e^{2\mu^{\ast}}(dr^{2}+r^{2} d\theta^{2}+r^{2}\sin^{2}{\theta} d\phi^{2})+e^{2\nu^{\ast}} dt^{2},$$
where
$$\begin{array}{llll}
  \nu^{\ast} &=&-\left[\dfrac{m_{1}}{r_{1}} + \dfrac{m_{2}}{r_{2}}\right]&=-\dfrac{m}{r}\\[10pt]
  \mu^{\ast} &=& \left[\dfrac{m_{1}}{r_{1}} + \dfrac{m_{2}}{r_{2}}\right]
  + \mathcal{O}\left(\dfrac{1}{r^{2}}\right)&=\dfrac{m}{r},
\end{array}$$
then the line element can be rewritten as
\begin{eqnarray*}
ds^{2}&=&-\left[1+\dfrac{m}{r}+\mathcal{O}\left(\dfrac{1}{r^{2}}\right)\right]^{2}
         (dr^{2} + r^{2}d\theta^{2} + r^{2} \sin^{2} \theta d\phi)\\[5pt]
      & &+\left[1+\dfrac{m}{r}+\mathcal{O}\left(\dfrac{1}{r^{2}}\right)\right]^{-2}
         dt^{2}.
\end{eqnarray*}
Apply the transformation $R=r+m$, we get
$$ds^{2}=-\left(1-\dfrac{2m}{R}\right)^{-1} dR^{2} - R^{2} d\theta^{2}-
R^{2} \sin^{2} \theta d\phi^{2} + \left(1-\dfrac{2m}{R}\right)
dt^{2},$$ which represents the line element of Schwarzschild
gravitational field in its standard form.
\subsection{Scalar invariant and Singularities}\label{Singularities}
according to the original solution given by Curzon, the solution is singular at the two masses of the binary system. We examine the solution in its modified version, after rotation, to check if the solution is free from new singularities or not. This can be done by checking the curvature scalar. Now the suggested singularities are raised at
\begin{eqnarray*}\label{invars}
\begin{split}
2e^{\mu}\left[\cos^{2}(\phi+\omega t)+\sin^{2}(\phi+\omega t)cos^{2}\theta \right]^{3}
\left[\cos^{2}\theta+\sin^{2}\theta \cos^{2}(\phi+\omega t)\right] r^{2}\sin^{2}{\theta}=0
\end{split}
\end{eqnarray*}
by solving the above equation for $\omega$, we find
\begin{eqnarray*}\label{sing1}
    \omega&=&\dfrac{-\phi \pm i~\tanh^{-1}\left({\csc{\theta}}\right)}{t},\\
    \omega&=&\dfrac{-\phi \pm \pi/2 \mp i~\sinh^{-1}\left({\cot{\theta}}\right)}{t},\\
\end{eqnarray*}
Taking initially $\theta=\pi/2$, we find the same solution for the above conditions
\begin{equation}\label{ssing1}
    \omega=\dfrac{-\phi \pm \pi/2}{t}.
\end{equation}
The effect of the rotation $\hat{\phi}=\phi+\omega t$ can be applied everywhere in the spacetime without need to a regularity condition. And so the solution is free from new singularities due to phase shift $\pi/2t$. In another word, one can say that the world line of the pulsar and its companion is characterized by
\begin{equation}\label{ssing1}
    t=\dfrac{-\phi \pm \pi/2}{\omega},
\end{equation}
where the (+ve) sign corresponds to corotation $``$pulsar" and the (-ve) one to counter-rotation $``$companion" half loop for each.
\subsection{Killing Vectors}\label{Killing}
In the general theory of relativity, no hope to solve Einstein's field equations exactly without imposing a symmetry. The axially
symmetric assumption is a very natural assumption. Geometrically, axial symmetry means the existence of a spacelike rotational
killing vector ${\partial}/{\partial \varphi}$. Also, we wish this spacetime to be at least locally asymptotically flat to describe finite sources. In addition, it appears hopeless to search for a radiative spacetime with only one symmetry. We summarize this in the following conditions
\begin{enumerate}
    \item A natural assumption is axially symmetric.
    \item The spacetime should be, at least, locally asymptotically flat.
\end{enumerate}
To gain better insight into curved spacetimes with a rotational symmetry let us first consider the Minkowski spacetime where the two Killing vectors and their norms have the form\\
\begin{enumerate}
    \item [$\bullet$] the axial Killing vector
\begin{equation}\label{axialkv}
    \xi=-\cos{(\phi+\omega t)}\partial_{\theta}+\cot{\theta}\sin{(\phi+\omega t)}\partial_{\phi},
\end{equation}
    \item [$\bullet$] the rotation Killing vector
\begin{equation}\label{boostkv}
    \eta=\partial_{t}-\omega \partial_{\phi}.
\end{equation}
\end{enumerate}

In fact, the whole structure of group orbits in rotation symmetric curved spacetimes is similar to the generated structure by the axial and rotation Killing vectors in Minkowiski space. One obtains a similar picture for a curved spacetimes with a rotational symmetry by checking the invariance of a metric (or of any other field) in a time direction.
\begin{subequations}
\renewcommand{\theequation}{\theparentequation.\arabic{equation}}
    \begin{eqnarray}
\xi_{\alpha}~ \nu&=& \cos{(\phi+\omega t)}\partial_{\theta}\nu-\cot{\theta} \sin{(\phi+\omega t)}\partial_{\phi} \nu=0,\label{k1} \\
\xi_{\alpha}~ \mu&=& \cos{(\phi+\omega t)}\partial_{\theta}\mu-\cot{\theta} \sin{(\phi+\omega t)}\partial_{\phi} \mu=0, \label{k2}\\
\eta_{\alpha}~ \nu&= & \omega \partial_{\phi} \nu - \partial_{t}\nu=0, \label{k3}\\
\eta_{\alpha}~ \mu&= & \omega \partial_{\phi} \mu - \partial_{t}\mu=0.\label{k4}
    \end{eqnarray}
\end{subequations}
Where $\mu$ and $\nu$ are the two functions characterizing the metric.\\

In Bonnor's work the transformation has been done between the radial coordinate $z$ and the temporal $t$, which gives a radiative property on the axis of symmetry as the two singularities vibrate \cite{Bon92, BGM94}. Actually the radiative properties are due to exchange projections (partially) of phenomenon between temporal and spacial coordinates. Here in the present work we expect the same partial exchange case, but between the azimuthal angle $\phi$ and the temporal $t$. Similarly, we expect a different quantitatively radiative behavior but more physical\footnote{This work is in progress now}.
\section{Motion in The Modified Curzon Field}\label{Motion in The Modified Curzon Field}
$``$\textit{spacetime tells matter how to move, and matter tells spacetime how to curve}$"$ \cite{MTW73}. We have seen how matter
tells spacetime how to curve, now we would like to search how spacetime tells matter how to move!! So we calculate the
non-vanishing Christoffel symbol coefficients of the second kind (symmetric in first two indices) for the space represented by
(\ref{q3.11.1})-(\ref{q3.11.7}). By using the calculated values of the Christoffel symbols and apply the geodesic equation,
$$\dfrac{d^{2}x^{\alpha}}{ds^{2}}+\Gamma^{\alpha}_{~\mu \nu}\dfrac{dx^{\mu}}{ds}\dfrac{dx^{\nu}}{ds}=0,$$
one can formulate the equations of motion of a test particle in the gravitational field of a binary system. In this way we can treat the motion of a test particle like a third body in the binary system, without adopting perturbation techniques. If so, we can describe the motion of a massless particle (e.g. photon), in the field of the binary pulsar using a null geodesic. To test planer motion, we put $x^{2}=\theta$ in the equations of motion and using the calculated values of the Christoffel symbols $\Gamma$, we can see that the differential equation for the angle $\theta$ can be written as,
$$\dfrac{d^{2}\theta}{ds^{2}} + \left[\mathcal{A} \dfrac{dr}{ds} + \mathcal{B}
\dfrac{d\theta}{ds} + \mathcal{C} \dfrac{d\phi}{ds} + \mathcal{D}
\dfrac{dt}{ds}\right]\dfrac{d\theta}{ds}=0,$$
where $\mathcal{A}$, $\mathcal{B}$, $\mathcal{C}$, and $\mathcal{D}$
are known functions. By taking initially
$\theta=\theta_{0}=\frac{\pi}{2}$ , and
$\left(\dfrac{d\theta}{ds}\right)_{0}=0$, we get from the above
equation
$$\dfrac{d^{2}\theta}{ds^{2}}=0,$$
then
\begin{equation}\label{thetadot}
\dot{\theta} \equiv \dfrac{d\theta}{ds}=0.
\end{equation}
It is clear from this solution that the motion of a test particle is
a planer motion. Now restricting ourselves by taking the plane
$\theta=\pi/2$ for an eclipsing binary. In this case the line
element will become
\begin{equation}\label{metric}
ds^{2}=-e^{\mu}dr^{2}-r^{2}e^{\mu} d\phi^{2} -2\omega r^{2}
e^{\mu}d\phi dt + \left(e^{\nu}-\omega^{2}r^{2}e^{\mu}\right)dt^{2}.
\end{equation}
For $x^{3}=\phi$, the equation of motion becomes,
\begin{equation}
\dfrac{d}{ds}\left[-2r^{2}e^{\mu}\dot{\phi}-2\omega r^{2} e^{\mu}
\dot{t}\right]=e^{\nu}\left(\nu_{\phi}-\mu_{\phi}\right)\dot{t}^{2},
\end{equation}
where $\mu_{\phi}=\dfrac{\partial \mu}{\partial \phi}$, and
$\nu_{\phi}=\dfrac{\partial \nu}{\partial \phi}$. This leads to
\begin{equation}\label{3.35}
r^{2}e^{\mu}\left(\dot{\phi}+\omega \dot{t}\right)=
-\dfrac{1}{2}\int e^{\nu}\left(\nu_{\phi}-\mu_{\phi}\right)\dot{t}~
dt + \mathscr{L},
\end{equation}
where $\mathscr{L}$ is an arbitrary constant.\\
Similarly, for $x^{4}=t$, we get
\begin{equation}
\dfrac{d}{ds}\left[-2\omega r^{2} e^{\mu} \dot{\phi}+
2\left(e^{\nu}-\omega^{2} r^{2} e^{\mu}\right)\dot{t}\right]=
e^{\nu}\left(\nu_{t}-\mu_{t}\right)\dot{t}^{2}.
\end{equation}
The above differential equation leads to
\begin{equation}\label{3.37}
\omega r^{2} e^{\mu}\left(\dot{\phi}+\omega \dot{t}\right)-e^{\nu}
\dot{t}= -\dfrac{1}{2} \int e^{\nu}\left(\nu_{t}-\mu_{t}\right)
\dot{t}~ dt-\mathscr{E},
\end{equation}
where $\mathscr{E}$ is another arbitrary constant.\\

Recalling equations (\ref{q3.12.1}), (\ref{q3.12.2}), (\ref{k3}) and (\ref{k4}), multiplying (\ref{3.35}) by the constant $\omega$, combining it with (\ref{3.37}) and solving the equations for $\dot{t}$, we get,
\begin{equation}\label{tdot}
\dot{t}=\left(\mathscr{E}+\omega \mathscr{L}\right)e^{-\nu}.
\end{equation}
Similarly, solving for $\dot{\phi}$, we get
\begin{equation}\label{phidot}
\dot{\phi}=\dfrac{\mathscr{L}}{r^{2}}e^{-\mu}- \omega
\left(\mathscr{E}+\omega
\mathscr{L}\right)e^{-\nu}-\left(\mathscr{E}+\omega
\mathscr{L}\right) F_{\phi},
\end{equation}
where
$$F_{\phi}=\dfrac{e^{-\mu}}{2r^{2}}\int \left(\nu_{\phi}-\mu_{\phi}\right)dt.$$
It is easy to show that $F_{\phi} \approx
\mathcal{O}\left(\dfrac{1}{r^{3}}\right)$, which can be ignored for
the time being, therefore
\begin{equation}\label{phidot3}
\dot{\phi} \approx \dfrac{\mathscr{L}}{r^{2}}e^{-\mu}-\omega
\left(\mathscr{E}+\omega
\mathscr{L}\right)e^{-\nu}+\mathcal{O}\left(\dfrac{1}{r^{3}}\right).
\end{equation}
Recalling equation (\ref{metric}), and substituting from
(\ref{tdot}) and (\ref{phidot3}), we get for $\dot{r}$
\begin{equation}\label{rdot}
\dot{r}=\sqrt{\left(\mathscr{E}+\omega
\mathscr{L}\right)^{2}e^{-\mu-\nu}-e^{-\mu}\left(B+\dfrac{\mathscr{L}^{2}}{r^{2}}e^{-\mu}\right)},
\end{equation}
where the parameter $B$ is defined as\\
$B=\left\{%
\begin{array}{ll}
    0, & \hbox{for a photon;} \\
    1, & \hbox{for a material test particle.} \\
\end{array}%
\right. $\\

In what follows, we are going to extract some physical features from
the above results by considering some special cases.
\subsection{Boundary Conditions on Equations of Motion}\label{The First Check: Flat Space Metric}
\smallskip
\subsubsection*{(a) The Linear Approximation Limiting Case:}
Assuming that the source of field is too distant from the observer, as done in section \ref{Check of the new solution}, then we
can consider that $r \simeq r_{1}\simeq r_{2}$. Also the total mass of the system is $m=m_{1}+m_{2}$. By taking the approximation $r \gg m$ so that $\mathcal{O}\left(\dfrac{m^{2}}{r^{2}}\right) \to 0$, then
$$e^{\mu}=e^{-\nu} \approx \left(1-\dfrac{2m}{r}\right)^{-1},$$
and
\begin{subequations}\label{3.38}
\renewcommand{\theequation}{\theparentequation.\arabic{equation}}
   \begin{eqnarray}
     \dot{\theta} &= & 0, \\[5pt]
     \dot{t} &\approx& \dfrac{\left(\mathscr{E}+\omega
     \mathscr{L}\right)}{\left(1-\dfrac{2m}{r}\right)},
     \\[5pt]
     \dot{\phi} &\approx& \dfrac{\mathscr{L}}{r^{2}}-\omega \dfrac{\left(\mathscr{E}+\omega \mathscr{L}\right)}
                           {\left(1-\dfrac{2m}{r}\right)}, \\[5pt]
     \dot{r} &\approx& \sqrt{\left(\mathscr{E}+\omega
                           \mathscr{L}\right)^{2}-\left(1-\dfrac{2m}{r}\right)
                           \left(B+\dfrac{\mathscr{L}^{2}}{r^{2}}\right)}.
   \end{eqnarray}
\end{subequations}
It is clear that the above equations are similar to the motion in the Schwarzschild case except for some additional terms depending on the angular velocity $\omega$. Also, from our experience in classical mechanics it is clear that the constants of integrations $\mathscr{L}$ and $\mathscr{E}$ are, respectively, representing the angular momentum and the energy of the moving test particle.
\smallskip
\subsubsection*{(b) The Static Field:} Taking $\omega = 0$, we get
the following set of differential equations:
\begin{subequations}\label{3.39}
\renewcommand{\theequation}{\theparentequation.\arabic{equation}}
   \begin{eqnarray}
     \dot{\theta} &= & 0, \\[5pt]
     \dot{t} &\approx&
     \dfrac{\mathscr{E}}{\left(1-\dfrac{2m}{r}\right)},
     \\[5pt]
     \dot{\phi} &\approx& \dfrac{\mathscr{L}}{r^{2}}, \\[5pt]
     \dot{r} &\approx& \sqrt{\mathscr{E}^{2}-\left(1-\dfrac{2m}{r}\right)
                           \left(B+\dfrac{\mathscr{L}^{2}}{r^{2}}\right)}.
   \end{eqnarray}
\end{subequations}
which is now identical to the motion in Schwarzschild field.\\

After all, our attempt is to show how some measurable quantities (e.g. redshift and time delay in the field of binary systems) can be extracted by using of the solution of the equations of motion.
\section{Time Delay in Binary Systems}\label{Time delay in binary systems}
Many measurable quantities can be evaluated by using the model suggested in the present work. One of these quantities is the time delay due to the gravitational field of a binary pulsar. Since the discovery of the first pulsar in a binary system \cite{HT75}, many trials have been done to describe the characteristics of the binary pulsar system in the frame of the general relativity, which gives good prediction compatible with the observational results, especially the orbital decay resulting from the emission of the gravitational radiation \cite{Wag75, WW76, Will77, TW82}. The most significant theory-independent models are given in the literature \cite{EH75, BT75, BT76, Eps77, Hgn85, DD85, DD86}, all of them are based on the arrival times of the pulses from a pulsar.\\

Also, many articles have been devoted to predict the time delay of the pulsation due to presence of the pulsar in the field of its companion \cite{THF76}. Furthermore, some authors have studied the effect of the rotation of the companion on the pulsation arrival times of the primary\cite{LW97}. Moreover, the effect of the gravito-magnetic correction on the Shapiro time delay due to the intrinsic angular momentum of the stars has been studied \cite{TRN05, RT05}.\\

\begin{figure}
\begin{center}
\includegraphics[scale=0.8]{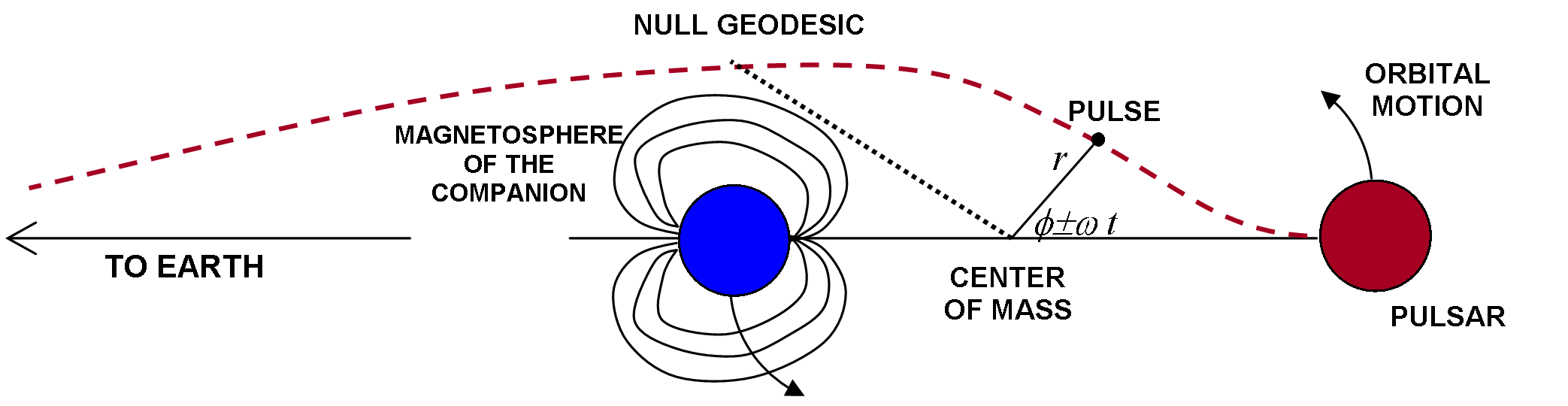}
\caption{the null path of pulses at the superior conjunction. It is clear that the contribution of the position of the pulsar and its companion to the time delay within the orbit size should be taken into account as expected for time dependent field.}
\label{fig1}
\end{center}
\end{figure}

The general relativity involves many post newtonian effects which still can be considered as relativistic gravitoelectric effects, c.f. \cite{I07}. Another effect can be studied in the present modified version of Curzon solution is the gravito-magnetic field due to the orbital motion of the pulsar and its companion. The gravito-magnetic clock effect involves a certain characteristic temporal structure around rotating source of gravitational field. But these effects are generally small to the so called gravito-electric interaction. This makes their detection very difficult. Many authors has investigated the gravito-magnetic effects in binary pulsar systems due to the rotation of the companion. They pointed out the difficulties of the detection of these effects \cite{DK95, LW97,WK99, KM02}, \cite{RL06a, RL06b,FB09, RT07, RT02}. In particular, we investigate other possible relativistic effects, in addition to original Shapiro time delay, resulting from the angular orbital motion of the members of binary system (gravito-magnetic) in a time dependent gravitational field. We have assumed that light rays (\textit{pulses}) are traveling from the pulsar of the binary system to the earth along a null-geodesic, which lies in the orbital plane ($\theta=\pi/2$). Also, we assume that its closest approach to the companion is
\begin{equation}\label{sep}
r \sin (\phi + \omega t) \approx d,
\end{equation}
Due to the strong curvature of the gravitational field we take $d$ as the radius of the magnetosphere of the companion rather than the impact parameter, see Figure (\ref{fig1}).
It is possible to obtain a relation between the time of arrival of a pulse at the Earth $t_{arr}$ and its time of emission $t_{em}$. This can be done by determining the relationship between the time and space coordinates along the world-line of a light ray. Writing the line element for an eclipsing binary system (\ref{metric}), after some easily algebraic calculations. we get the line element for a ray of light,
\begin{equation}\label{m0}
0=-e^{\mu}dr^{2}-r^{2}e^{\mu}\left(d \phi+\omega d
t\right)^{2}+e^{\nu}dt^{2}.
\end{equation}
From (\ref{sep}) we can reexpress $r^{2} \left(d \phi+\omega d
t\right)^{2}$ in terms of $r$ and $dr$, so that
\begin{eqnarray}\label{dt}
dt^{2}&=&e^{\mu-\nu}\sec^{2}\left(\phi+\omega t\right) dr^{2},\notag\\
      &=&\dfrac{e^{\mu-\nu}dr^{2}}{\left(1-d^{2}/r^{2}\right)}.
\end{eqnarray}
In the above equation the potential functions (\ref{q3.12.1}),
(\ref{q3.12.2}) are written, in terms of the radial coordinate $r$,
as
\[
\nu  =  - {\displaystyle \frac {2\,{m_{1}}}{\sqrt{(r + a)^{2} -
2\,a\,r\,(1 + {\displaystyle {d/r}} )}}}  - {\displaystyle
\frac {2\,{m_{2}}}{\sqrt{(r + a)^{2} - 2\,a\,r\,(1
 - {\displaystyle {d/r}} )}}},
\]
\noindent and
\begin{eqnarray}
  \mu & = & {\displaystyle \frac {2\,{m_{1}}}{\sqrt{(r + a)^{2} - 2\, a\,r\,(1 + {\displaystyle {d/r}} )}}}  +
            {\displaystyle \frac {2\,{m_{2}}}{\sqrt{(r + a)^{2} - 2\,a\,r\,(1 - {\displaystyle {d/r}} )}}} \notag\\ [10pt]
      & - & r^{2}\,\left(1 - {\displaystyle \frac {d^{2}}{r^{2}}} \right)\,\left[  \! {\displaystyle \frac {{m_{1}}^{2}}{\left[(r + a)^{2} - 2\,a\,r\,(1 + {\displaystyle {d/r}} )\right]^{2}}}  + {\displaystyle \frac {{m_{2}}^{2}}{\left[(r + a)^{2} - 2\,a\,r\,(1 -
            {\displaystyle {d/r}} )\right]^{2}}}  \!  \right]\notag\\ [10pt]
      & + & {\displaystyle \frac {{m_{1}}\,{m_{2}}}{a^{2}}\, \left[  \!
            {\displaystyle \frac {r^{2} - a^{2}}{\sqrt{(r + a)^{2} - 2\,a\,r
            \,(1 + {\displaystyle {d/r}} )}\,\sqrt{(r + a)^{2} - 2\,a
            \,r\,(1 - {\displaystyle {d/r}} )}}}  - 1 \!  \right] }.
\end{eqnarray}
\noindent We now take the square root of (\ref{dt}), expanding $dt$
to the order $\mathcal{O}(r^{-4})$, then integrating as usual, we get the total time $\delta t$
\begin{equation*}
\delta t= \int_{t_{em}}^{t_{arr}} dt=\int_{t_{em}}^{t_{d}} dt+\int_{t_{d}}^{t_{arr}} dt.
\end{equation*}
We obtain the time of flight required for a pulse to travel, in the equatorial plane, from the binary system to an arbitrary point $r$, in the form
\begin{equation}\label{tdd}
\begin{split}
\delta t_{\pm}&= \sqrt{r^2-d^2}-\sqrt{a^2-d^2}+2(m_{1}+m_{2})\ln{\left(\frac{r+\sqrt{r^2-d^2}}{a+\sqrt{a^2-d^2}}\right)}+a(m_{1}+m_{2})\left[\frac{r\sqrt{a^2-d^2}-a\sqrt{r^2-d^2}}{r~d^2}\right] \\[10pt]
&+\frac{3(m_{1}+m_{2})^2}{2d}\left[\arctan\left(\frac{d(\sqrt{r^2-d^2}-a)}{\sqrt{r^2-d^2}+d^2}\right)-\arctan\left(\frac{d(\sqrt{a^2-d^2}-a)}{\sqrt{a^2-d^2}+d^2}\right)\right] \\[10pt]
&-\frac{\left(m_{1}+m_{2}\right)^{3}}{3}\left[\frac{r\sqrt{a^2-d^2}-a\sqrt{r^2-d^2}}{a~r~d^2}\right]\mp 2(m_{2}-m_{1})\left[\frac{r\sqrt{a^2-d^2}-a\sqrt{r^2-d^2}}{r~d}\right].
\end{split}
\end{equation}
\noindent The above equation gives an explicit form of the time delay of the pulses due to time dependent gravitational field of a binary system. Let $\delta t_{\pm}$ be the time of flight when revolution is in the same (opposite) sense as the rotation of the source. It shows that the time of flight dependance on the masses of the two members of the binary system $m_{1}$ and $m_{2}$. This can be easily understood since the gravitational field in the suggested model is due to the field of the two masses of the binary system, which is different from case of the Schwarzschild field. The first two terms in the above equation corresponds to signal propagation in flat space-time. The third term is the original gravitational Shapiro time delay. While the fourth term is the geometrical time delay, usually discarded in the Solar system. The fifth term is a third order correction in masses to Shapiro time delay. In particular we are interested in the gravito-magnetic term which is presented by the last term. This type of delay is sensitive to the change of the phase of the two members of the binary system. Its max effect is near the superior conjunction. Since it is antisymmetric, its contribution can be separated from other time delays. This can be done by taking the time difference between two identical signals emitted from the system when it rotates in two different directions.
\begin{equation}\label{timediff}
\begin{split}
    \Delta t=&\left[\delta t_{+}(r_{em},r_{d})+\delta t_{+}(r_{d},r_{arr})\right]-\left[\delta t_{-}(r_{em},r_{d})+\delta t_{-}(r_{d},r_{arr})\right],\\
    =&4(m_{1}-m_{2})\left[\frac{r\sqrt{a^2-d^2}-a\sqrt{r^2-d^2}}{r~d}\right].
\end{split}
\end{equation}
Actually the typical radius of a neutron star is about 10 km which makes a large delay to Shapiro time. Unfortunately, the effect of a large magnetosheath keeps the magnitude of the gravito-magnetic correction not that large as expected, according to equation (\ref{timediff}), few $\mu s$ $\approx 0.018\pm0.001$ $\mu s$, as pulses are not received during the period of the eclipse by the companion magnetosheath. Although The smallness of the gravito-magnetic delay would require long data taking times, it is not affected by any drift or noise. So it should emerged in the long period phenomena. While we expect a significant value of this effect of this delay for binary systems during their coalescence.\\

One can see an additional terms in equation (\ref{tdd}) due a shift of the coordinate system to the center of mass of the binary system. In order to compare our result to previous treatments, we shift the frame of reference to be at the center of the companion, and reexpress the obtained equation (\ref{tdd}) we get
\begin{equation}\label{shtd}
\begin{split}
\delta t_{\pm}=&\sqrt{r^2-d^2}+2\left(m_{1}+m_{2}\right)\ln\left({\frac{r+\sqrt{r^2-d^2}}{d}}\right)-\left[\left(m_{1}+m_{2}\right)a^{2} \mp 2ad\left(m_{1}-m_{2}\right)\right]\frac{\sqrt{r^{2}-d^{2}}}{d^{2} r}\\[6pt]
&+\frac{3}{4}\frac{\left(m_{1}+m_{2}\right)^{2}}{d}\left[\pi-2 \arctan{\left(\frac{d}{\sqrt{r^{2}-d^{2}}}\right)}\right]+\frac{1}{3}\left(m_{1}+m_{2}\right)^{3}\frac{\sqrt{r^{2}-d^{2}}}{d^{2} r}.
\end{split}
\end{equation}
The above equation is very similar to the time delay in kerr field when the spin of the companion is taken into account and time of flight within the system is ignored, see \cite{D86,LW97}. The difference from the work of \cite{D86} is that the last term here represent the gravito-magnetic effect due to the orbital motion of the pulsar and its companion not the spin of the companion.
\subsection{Applications to PSR J0737-3039A \& B}
\begin{figure}
\begin{center}
\includegraphics[scale=0.2]{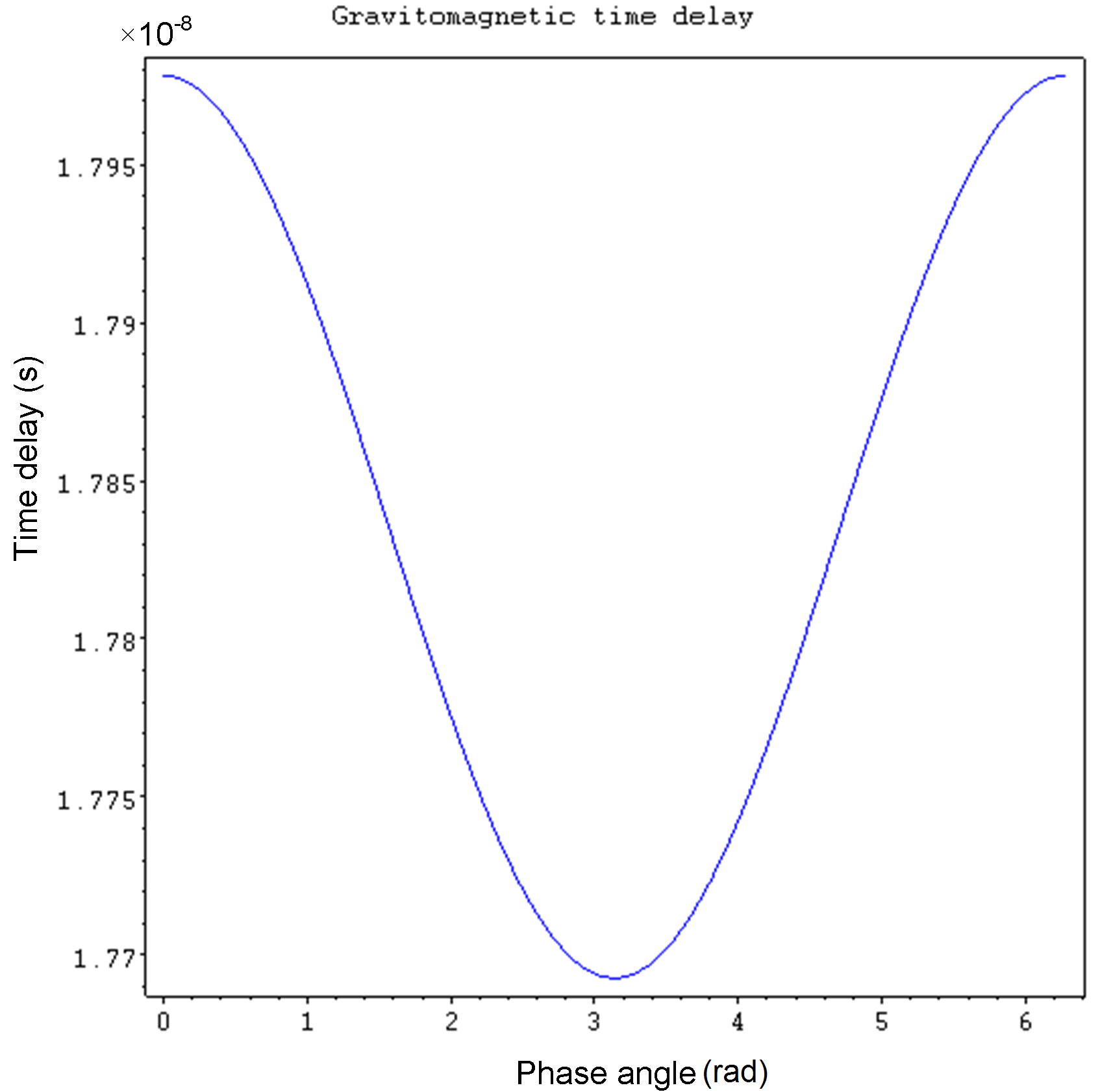}
\caption{A model curve for the gravito-magnetic time delay over one orbital revolution. Its maximum effect at the superior conjunction. This effect is accumulative so it requires a long time observations to be taken into account.}
\label{fig2}
\end{center}
\end{figure}

We next apply the obtained equation for the time of flight (\ref{tdd}) on the double pulsar system PSR J0737-3039. This system is the only system observed in which both members are neutron stars, so far, with comparable masses which are pulsating. It has a nearly circular orbit with an eccentricity $e$ = 0.088 and angle of inclination is 88.69$^{\circ}$ so it is observed nearly edge-on. Also this binary has a radius of 1.25$R_{\odot}$, thus the entire system could fit within our sun. Also it has a very small orbital period of 2.45 hours allowing a rapid accumulating relativistic effects of higher orders. Other useful information appear in Table (1).
\begin{center}
\textbf{Table 1}: Some observed quantities of the binary pulsar PSR J0737-3039 A$\&$B \cite{Bur03, Lyne04}.
\end{center}
\begin{center}
\begin{tabular}{ccccc}
\hline
\hline
distance     & orbital radius & angular velocity & Mass A  & Mass B  \\
 $r$ (lightyear) & $a$ (km)         & $\omega$ ($s^{-1}$)        & $M_{2}$ &  $M_{1}$\\
\hline
 1700          & 1.25$R_{\odot}$& $7\times10^{-4}$         & 1.25(5)$M_{\odot}$ & 1.37(5)$M_{\odot}$\\
\hline
\end{tabular}
\end{center}
For this system we give an estimated numerical values of the contribution of each term according to equation (\ref{tdd}). We use the relativistic units $ c = G = 1$. Assuming that the closest approach to the companion ($\approx 1.81 \times 10^4$ km) at the conjunction line as seen during the eclipse observations c.f. \cite{K04}. We get the first and second terms are of an order of magnitude $0.5364791999\times10^{11}$ s $\approx$ 1700 lightyear, which represents the time of flight in flat space from the system to Earth. The contribution of the mass to the (geometric) time delay is $6.38\pm0.013$ $\mu s$. The max Shapiro time delay is $602.8\pm1.25$ $\mu s$ due to presence of the field of both members $M_{1}$ and $M_{2}$ at the closest approach to the companion at the conjunction line. The second order correction to the mass term is $8.3\pm0.035\times10^{-9}$ s. The third order correction to the mass term is $4.1\pm0.026\times10^{-15}$ s. The contribution of the gravito-magnetic term is of an order of magnitude $0.89\pm0.05\times10^{-8}$ s. Also we draw a model curve for for equation (\ref{timediff}) the gravito-magnetic time delay during one revolution, see Figure (\ref{fig2}). This correction and similar corrections, c.f. \cite{LR05}, should be taking into account especially during the coalescence phase. As we mention before this term is not affected by a drift or noise and it requires long time observation. However, this effect will be affective during the last phase of the evolution of the system. It is expected to be $0.21\pm0.013 ~\mu s$ when the separation is nearly $0.22~R_{\odot}$.
\section{Conclusion and Prospective}\label{Prospective}
In the present work, we discussed two major problems the two body problem in the general theory of relativity and the effect of the rotating frame of reference on the spacetime. The first problem has been tackled before by Curzon, but the static configuration of the solution he obtained does not give any physical representation in Nature. Also, we mentioned that the trials to generate stationary solutions, to give what is like double-Kerr solution in order to hold the two singularities apart, are also un-physical. Further, we visited another important problem which is the effect of a rotating (non-inertial) frame in the special relativity theory and the general relativity theory as well. In another word, on the curvature of spacetime geometry.\\

This led us to rewrite the Curzon solution in a rotating frame of reference suitable for describing the orbital motion of the two masses. This enabled us to give a new forsight on the physical quantities relative to an observer at rest in the rotating frame. We checked that the solution does not need any additional regularity condition to keep the rotation on. Also we give the killing vector due to rotational symmetry imposed by rotation. Moreover, we derived and solved the equations of motion in the modified version of the Curzon field. Furthermore, we used the obtained equations to calculate the time delay in the binary pulsar systems. Finally we applied the obtained results in the case of the binary pulsars PSR J0737-3039A \& B. The study of the gravito-magnetic effects on the gravitational time delay appears as a correction to the gravito-electric contribution. These effects are generally small, since the coupling with the orbital angular momentum of the binary is much weaker than the coupling with mass alone. This makes their detection difficult. While the gravito-magnetic interaction has a leading role when studying the phenomenon of the gravitational Faraday rotation \cite{RT07}.\\

The authors also would like to mention that the modified solution enables them to go over to a more useful benefit of this model which is its capability to calculate the redshift, on a curved spacetime, of the pulsations of highly relativistic systems (binary pulsars) PSR J0737-3039A \& B like, in a natural way. Also, The model enables them to calculate the energy of the binary pulsars by using one of the famous methods (e.g. M\"{o}ller energy-momentum complex) c.f. \cite{Mlr58, MWLH96}. In this way we can estimate the decay of the orbital period due to gravitational radiation, coalescence rate and the gravitational radiation of the compact binary systems when it is not necessary to require the weak-field initial data. Furthermore, the model will be efficient to study a triple pulsar system PSR B1620-26 \cite{TACL99}.
\bibliographystyle{unsrt}
\bibliography{binary_v5}

\begin{thebibliography}{10}

\bibitem{Czn24}
H.~E.~J. Curzon.
\newblock Cylinderical solutions of {E}instein's gravitation equations.
\newblock {\em Proc. London Math. Soc.}, 23:477--480, 1924.

\bibitem{LB01}
Luc Blanchet.
\newblock On the two-body problem in general relativity.
\newblock {\em C. R. Acad. Sci. Paris, t. 2, S\'{e}rie IV}, pages 1343--1352,
  July 2001.

\bibitem{DD81}
Thibaut Damour and Nathalie Deruelle.
\newblock Radiation reaction and angular momentum loss in small angle
  gravitational scattering.
\newblock {\em Physics Letter}, 87A(3):81--84, December 1981.

\bibitem{Will93}
Clifford~M. Will.
\newblock {\em Theory and experiments in gravitational physics}.
\newblock Cambridge University Press, Cambridge, revised edition, 1993.

\bibitem{KSMH80}
D.~Kramer, H.~Stephani, M.~Mac{C}allum, and E.~Herlt.
\newblock {\em Exact Solutions of {E}instein's Field Equations}.
\newblock Cambridge University Press, Cambridge, 1980.

\bibitem{Bondi97}
H.~Bondi and J.~Samuel.
\newblock {The Lense-Thirring effect and Mach's principle}.
\newblock {\em Physics Letters A}, 228:121--126, 1997.
\newblock \url{http://adsabs.harvard.edu/abs/1997PhLA..228..121B}.

\bibitem{Bur03}
M.~Burgay, N.~D'Amico, A.~Possenti, R.N. Manchester, A.G. Lyne, B.C. Joshi,
  M.A. McLaughlin, M.~Kramer, J.M. Sarkisian, F.~Camilo, V.~Kalogera, C.~Kim,
  and D.R. Lorimer.
\newblock An increased estimate of the merger rate of double neutron stars from
  observations of a highly relativistic system.
\newblock {\em Nature}, 426:531--533, Dec 2003.

\bibitem{Lyne04}
A.~G. {Lyne}, M.~{Burgay}, M.~{Kramer}, A.~{Possenti}, R.~N. {Manchester},
  F.~{Camilo}, M.~A. {McLaughlin}, D.~R. {Lorimer}, N.~{D'Amico}, B.~C.
  {Joshi}, J.~{Reynolds}, and P.~C.~C. {Freire}.
\newblock {A Double-Pulsar System: A Rare Laboratory for Relativistic Gravity
  and Plasma Physics}.
\newblock {\em Science}, 303:1153--1157, February 2004.
\newblock astro-ph/0401086.

\bibitem{D10}
D.~{Dieks}.
\newblock {Space, Time and Coordinates in a Rotating World}.
\newblock {\em ArXiv e-prints}, January 2010.
\newblock \url{http://adsabs.harvard.edu/abs/2010arXiv1002.0130D}.

\bibitem{K03}
Robert~D. Klauber.
\newblock Derivation of the general case sagnac result using
  non-time-orthogonal analysis.
\newblock {\em Foundations of Physics Letters}, 16:447--463, 2003.
\newblock 10.1023/B:FOPL.0000012776.04871.6d.

\bibitem{RR02}
G.~Rizzi and M.~L. Ruggiero.
\newblock Space geometry of rotating platforms: an operational approach.
\newblock {\em Found.Phys.}, 32:1525--1556, 2002.
\newblock \url{http://adsabs.harvard.edu/abs/2002gr.qc.....7104R}.

\bibitem{C07}
H~Culetu.
\newblock Does rotation generate a massive string?
\newblock {\em Journal of Physics: Conference Series}, 66(1):012055, 2007.

\bibitem{JF04}
J.F. Pascual-Sanchez.
\newblock {Speed of gravity and gravitomagnetism}.
\newblock {\em Int.J.Mod.Phys.}, D13:2345--2350, 2004.

\bibitem{Bon92}
W.~B. Bonnor.
\newblock Physical interpretation of vaccum solutions of einstein's equations.
  \textmd{{Part I}}. time-independent solutions.
\newblock {\em Gen. Rel. Grav.}, 24:551, 1992.

\bibitem{RN68}
H.~P. Robertson and Thomas~W. Noonan.
\newblock {\em Relativity and Cosmology}.
\newblock W. B. Saunders company, Philadelphia, 1968.

\bibitem{S36}
Ludwik Silberstein.
\newblock Two-centers solution of the gravitational field equations, and the
  need for a reformed theory of matter.
\newblock {\em Phys. Rev.}, 49:268--270, Feb 1936.

\bibitem{S85}
Nathan Schleifer.
\newblock Condition of elementary flatness and the two-particle curzon
  solution.
\newblock {\em Physics Letters A}, 112(5):204 -- 207, 1985.

\bibitem{SM73}
P.~Szekeres and F.~H. Morgan.
\newblock Extensions of the curzon metric.
\newblock {\em Communications in Mathematical Physics}, 32:313--318, 1973.
\newblock 10.1007/BF01645612.

\bibitem{ER36}
A.~Einstein and N.~Rosen.
\newblock Two-body problem in general relativity theory.
\newblock {\em Phys. Rev.}, 49:404--405, Mar 1936.

\bibitem{S68}
J.~Stachel.
\newblock Structure of the curzon metric.
\newblock {\em Physics Letters A}, 27(1):60 -- 61, 1968.

\bibitem{HCD84}
Cornelius Hoenselaers and Werner Dietz.
\newblock The rankn hkx transformations: New stationary axisymmetric
  gravitational fields.
\newblock {\em General Relativity and Gravitation}, 16:71--78, 1984.
\newblock 10.1007/BF00764018.

\bibitem{GS03}
T.~I. {Gutsunaev} and A.~A. {Shaideman}.
\newblock {Axially symmetric gravitational fields. III. New methods of
  generating stationary vacuum Einstein fields}.
\newblock {\em Gravitation and Cosmology}, 9:229--234, December 2003.

\bibitem{H90}
C.~{Hoenselaers}.
\newblock {Axisymmetric stationary vacuum solutions near a ring singularity}.
\newblock {\em Classical and Quantum Gravity}, 7:581--584, April 1990.

\bibitem{DH82}
W.~Dietz and C.~Hoenselaers.
\newblock Stationary system of two masses kept apart by their gravitational
  spin-spin interaction.
\newblock {\em Phys. Rev. Lett.}, 48:778--780, Mar 1982.

\bibitem{DH85}
Werner Dietz and C~Hoenselaers.
\newblock Two mass solutions of einstein's vacuum equations: The double kerr
  solution.
\newblock {\em Annals of Physics}, 165(2):319 -- 383, 1985.

\bibitem{BGM94}
W.~B. Bonnor, J.~B. Griffiths, and M.~A.~H. MacCallum.
\newblock Physical interpretation of vaccum solutions of einstein's equations.
  \textmd{Part II}. time-dependent solutions.
\newblock {\em Gen. Rel. Grav.}, 26:687, 1992.

\bibitem{MTW73}
C.~W. Misner, K.~S. Thorn, and J.~A. Wheeler.
\newblock {\em Gravitation}.
\newblock Freeman, San Francisco, 1973.

\bibitem{HT75}
R.~A. Hulse and J.~H. Taylor.
\newblock Discovery of a pulsar in a binary system.
\newblock {\em The Astrophysical Journal}, 195:L51--L53, January 1975.

\bibitem{Wag75}
R.~V. {Wagoner}.
\newblock {Test for the existence of gravitational radiation}.
\newblock {\em The Astrophysical Journal Letter}, 196:L63--L65, March 1975.
\newblock
  \url{http://adsabs.harvard.edu/cgi-bin/nph-bib_query?bibcode=1975ApJ...196L..63W&db_key=AST}.

\bibitem{WW76}
R.~V. {Wagoner} and C.~M. {Will}.
\newblock {Post-Newtonian gravitational radiation from orbiting point masses}.
\newblock {\em The Astrophysical Journal}, 210:764--775, December 1976.
\newblock
  \url{http://adsabs.harvard.edu/cgi-bin/nph-bib_query?bibcode=1976ApJ...210..764W&db_key=AST}.

\bibitem{Will77}
Clifford~M. Will.
\newblock {Gravitational radiation from binary systems in alternative metric
  theories of gravity - Dipole radiation and the binary pulsar}.
\newblock {\em The Astrophysical Journal}, 214:826--839, June 1977.
\newblock
  \url{http://adsabs.harvard.edu/cgi-bin/nph-bib_query?bibcode=1977ApJ...214..826W&db_key=AST}.

\bibitem{TW82}
J.~H. {Taylor} and J.~M. {Weisberg}.
\newblock {A new test of general relativity - Gravitational radiation and the
  binary pulsar PSR 1913+16}.
\newblock {\em The Astrophysical Journal}, 253:908--920, February 1982.
\newblock
  \url{http://adsabs.harvard.edu/cgi-bin/nph-bib_query?bibcode=1982ApJ...253..908T&db_key=AST}.

\bibitem{EH75}
L.~W. {Esposito} and E.~R. {Harrison}.
\newblock {Properties of the Hulse-Taylor binary pulsar system}.
\newblock {\em The Astrophysical Journal (Letters)}, 196:L1+, February 1975.
\newblock
  \url{http://adsabs.harvard.edu/cgi-bin/nph-bib_query?bibcode=1975ApJ...196L...1E&db_key=AST}.

\bibitem{BT75}
R.~Blandford and S.~A. Teukolosky.
\newblock On the measurment of the mass of {P}{S}{R} 1913+16.
\newblock {\em The Astrophysical Journal}, 198:L27--L29, May 1975.

\bibitem{BT76}
R.~Blandford and S.~A. Teukolosky.
\newblock Arrival-time analysis for a pulsar in a binary system.
\newblock {\em The Astrophysical Journal}, 205:580--591, April 1976.

\bibitem{Eps77}
Reuben Epstein.
\newblock The binary pulsar: post-newtonian timing effects.
\newblock {\em The Astrophysical Journal}, 216:92--100, August 1977.

\bibitem{Hgn85}
Mark~P. Haugan.
\newblock Post-newtonian arrival-time analysis for a pulsar in a binary system.
\newblock {\em The Astrophysical Journal}, 296:1--12, September 1985.

\bibitem{DD85}
T.~Damour and N.~Deruelle.
\newblock General relativistic celestial mechanics of binary systems {I}. the
  post-newtonian motion.
\newblock {\em Ann. Inst. Henri Poincar\'{e}-Physique th\'{e}orique},
  43(1):107--132, 1985.

\bibitem{DD86}
T.~Damour and N.~Deruelle.
\newblock General relativistic celestial mechanics of binary systems {I}{I}.
  the post-newtonian timing formula.
\newblock {\em Ann. Inst. Henri Poincar\'{e}-Physique th\'{e}orique},
  44(3):263--292, 1986.

\bibitem{THF76}
J.~H. Taylor, R.~A. Hulse, L.~A. Fowler, G.~E. Gullahorn, and J.~M. Rankin.
\newblock Further observations of the binary pulsar {P}{S}{R} 1913+16.
\newblock {\em The Astrophysical Journal}, 206:L53--L58, May 1976.

\bibitem{LW97}
P.~{Laguna} and A.~{Wolszczan}.
\newblock {Pulse Arrival Times from Binary Pulsars with Rotating Black Hole
  Companions}.
\newblock {\em Astrophysical Journal Letter}, 486:L27+, September 1997.
\newblock astro-ph/9705054.

\bibitem{TRN05}
Angelo Tartaglia, Matteo~Luca Ruggiero, and Alessandro Nagar.
\newblock Time delay in binary systems.
\newblock {\em Phys. Rev. D}, 71:023003, Jan 2005.

\bibitem{RT05}
M.~L. {Ruggiero} and A.~{Tartaglia}.
\newblock {Post-Keplerian parameter to test gravitomagnetic effects in binary
  pulsar systems}.
\newblock {\em Physics Rev. D}, 72(8):084030, October 2005.
\newblock \url{http://adsabs.harvard.edu/abs/2005PhRvD..72h4030R}.

\bibitem{I07}
L.~{Iorio}.
\newblock The post-newtonian mean anomaly advance as further post-keplerian
  parameter in pulsar binary systems.
\newblock {\em Astrophys Space Sci}, 312:331--335, December 2007.

\bibitem{DK95}
O.~V. {Doroshenko} and S.~M. {Kopeikin}.
\newblock Relativistic effect of gravitational deflection of light in binary
  pulsars.
\newblock {\em Monthly Notices of the Royal Astronomical Society},
  274:1029--1038, June 1995.
\newblock \url{http://adsabs.harvard.edu/abs/1995MNRAS.274.1029D}.

\bibitem{WK99}
N.~{Wex} and S.~M. {Kopeikin}.
\newblock {Frame Dragging and Other Precessional Effects in Black Hole Pulsar
  Binaries}.
\newblock {\em astrophysical Journal}, 514:388--401, March 1999.
\newblock \url{http://adsabs.harvard.edu/abs/1999ApJ...514..388W}.

\bibitem{KM02}
S.~{Kopeikin} and B.~{Mashhoon}.
\newblock {Gravitomagnetic effects in the propagation of electromagnetic waves
  in variable gravitational fields of arbitrary-moving and spinning bodies}.
\newblock {\em Physics Rev. D}, 65(6):064025, March 2002.
\newblock \url{http://adsabs.harvard.edu/abs/2002PhRvD..65f4025K}.

\bibitem{RL06a}
R.~R. {Rafikov} and D.~{Lai}.
\newblock {Effects of gravitational lensing and companion motion on the binary
  pulsar timing}.
\newblock {\em Physics Rev. D}, 73(6):063003, March 2006.
\newblock \url{http://adsabs.harvard.edu/abs/2006PhRvD..73f3003R}.

\bibitem{RL06b}
R.~R. {Rafikov} and D.~{Lai}.
\newblock {Effects of Pulsar Rotation on Timing Measurements of the Double
  Pulsar System J0737-3039}.
\newblock {\em astrophysical Journal}, 641:438--446, April 2006.
\newblock \url{http://adsabs.harvard.edu/abs/2006ApJ...641..438R}.

\bibitem{FB09}
C.~Feiler, M.~Buser, E.~Kajari, W.~Schleich, E.~Rasel, and R.~O'Connell.
\newblock New frontiers at the interface of general relativity and quantum
  optics.
\newblock {\em Space Science Reviews}, 148:123--147, 2009.
\newblock 10.1007/s11214-009-9613-7.

\bibitem{RT07}
Matteo~Luca Ruggiero and Angelo Tartaglia.
\newblock {Gravitational Faraday Rotation in Binary Pulsar Systems}.
\newblock {\em Mon.Not.Roy.Astron.Soc.}, 374:847--851, 2007.
\newblock \url{http://arxiv.org/abs/astro-ph/0609712v2}.

\bibitem{RT02}
M.~L. {Ruggiero} and A.~{Tartaglia}.
\newblock {Gravitomagnetic effects}.
\newblock {\em Nuovo Cimento B Serie}, 117:743, July 2002.
\newblock \url{http://adsabs.harvard.edu/abs/2002NCimB.117..743R}.

\bibitem{D86}
I.~G. {Dymnikova}.
\newblock {Gravitational time delay of signals in the Kerr metric}.
\newblock In J.~{Kovalevsky} and V.~A. {Brumberg}, editors, {\em IAU Symp. 114:
  Relativity in Celestial Mechanics and Astrometry. High Precision Dynamical
  Theories and Observational Verifications}, pages 411--416, 1986.
\newblock
  \url{http://adsabs.harvard.edu/cgi-bin/nph-bib_query?bibcode=1986IAUS..114..411D&db_key=AST}.

\bibitem{K04}
V.~M. Kaspi, S.~M. Ransom, D.~C. Backer, R.~Ramachandran, P.~Demorest,
  J.~Arons, and A.~Spitkovsky.
\newblock Green bank telescope observations of the eclipse of pulsar "a" in the
  double pulsar binary psr j0737–3039.
\newblock {\em The Astrophysical Journal Letters}, 613(2):L137, 2004.

\bibitem{LR05}
Dong Lai and Roman~R. Rafikov.
\newblock Effects of gravitational lensing in the double pulsar system
  j0737–3039.
\newblock {\em The Astrophysical Journal Letters}, 621(1):L41, 2005.

\bibitem{Mlr58}
C.~M{\o}ller.
\newblock On the localization of the energy of a physical system in the general
  theory of relativity.
\newblock {\em Ann. Phys. (NY)}, 4:347, 1958.

\bibitem{MWLH96}
F.~I. Mikhail, M.~I. Wanas, E.~I. Lashin, and Ahmed Hindawi.
\newblock Sphericaly symmetric solutions in m{\o}ller's tetrad theory of
  gravitation.
\newblock {\em Int. J. Theor. Phys.}, 32:1627, 1993.
\newblock \url{http://arXiv.org/abs/gr-qc/9409039}.

\bibitem{TACL99}
S.~E. {Thorsett}, Z.~{Arzoumanian}, F.~{Camilo}, and A.~G. {Lyne}.
\newblock {The Triple Pulsar System PSR B1620-26 in M4}.
\newblock {\em The Astrophysical Journal}, 523:763--770, October 1999.
\newblock astro-ph/9903227.

\end{thebibliography}
\end{document}